\documentclass[journal=jacsat,manuscript=article]{achemso}

\usepackage{chemformula} 
\usepackage[T1]{fontenc} 
\usepackage{graphicx}
\usepackage{subcaption} 
\usepackage{float}  
\usepackage{comment}

\usepackage{soul}
\usepackage{xcolor}

\sethlcolor{cyan}

\author{Md Mushfiqul Islam}
\affiliation[USF-CS]
{Bellini College of Artificial Intelligence, Cybersecurity \& Computing, University of South Florida, Tampa, Florida, U.S.A.}

\author{Nishat N. Labiba}
\affiliation[USF-CS]
{Bellini College of Artificial Intelligence, Cybersecurity \& Computing, University of South Florida, Tampa, Florida, U.S.A.}

\author{Lawrence O. Hall}
\email{lohall@usf.edu}
\affiliation[USF-CS]
{Bellini College of Artificial Intelligence, Cybersecurity \& Computing, University of South Florida, Tampa, Florida, U.S.A.}

\author{David S. Simmons}
\email{dssimmons@usf.edu}
\affiliation[USF-CBME]
{Department of Chemical, Biological, and Materials Engineering, University of South Florida, Tampa, Florida, U.S.A.}

\title[Predicting Surface Tension]
  {Reducing Data Requirements for Sequence-Property Prediction in Copolymer Compatibilizers via Deep Neural Network Tuning}

\abbreviations{IR,NMR,UV}

\begin{document}







\begin{abstract}
  Synthetic sequence-controlled polymers promise to transform polymer science by combining the chemical versatility of synthetic polymers with the precise sequence-mediated functionality of biological proteins. However, design of these materials has proven extraordinarily challenging, because they lack the massive datasets of closely related evolved molecules that accelerate design of proteins. Here we report on a new Artifical Intelligence strategy to dramatically reduce the amount of data necessary to accelerate these materials' design. We focus on data connecting the repeat-unit-sequence of a \emph{compatibilizer} molecule to its ability to reduce the interfacial tension between distinct polymer domains. The optimal sequence of these molecules, which are essential for applications such as mixed-waste polymer recycling, depends strongly on variables such as concentration and chemical details of the polymer. With current methods, this would demand an entirely distinct dataset to enable design at each condition. Here we show that a deep neural network trained on low-fidelity data for sequence/interfacial tension relations at one set of conditions can be rapidly tuned to make higher-fidelity predictions at a distinct set of conditions, requiring far less data that would ordinarily be needed. This priming-and-tuning approach should allow a single low-fidelity parent dataset to dramatically accelerate prediction and design in an entire constellation of related systems. In the long run, it may also provide an approach to bootstrapping quantitative atomistic design with AI insights from fast, coarse simulations. 
  
\end{abstract}

\section{Introduction}
One of the greatest emerging challenges in 21st century polymer science is the understanding and design of synthetic sequence-controlled copolymers.\cite{perry_100th_2020} Thanks to advances in sequence-controlled synthesis over the past two decades, these polymers can now exhibit levels of repeat-unit sequence control rivaling that of biological macromolecules such as proteins and DNA.\cite{badi_sequence_2009,lutz_sequence-controlled_2013, lutz_future_2023, rosales_polypeptoids:_2013} However, unlike in biomacromolecules, design of these molecules cannot rely on data emanating from nearly 4 billion years of natural sequence evolution. Indeed, the central modern AI-based tools for protein structure prediction and design rely upon reasonable similarity of the designed molecule to naturally occurring proteins.\cite{abramson_accurate_2024,jumper_highly_2021} When designing synthetic molecules involving entirely distinct non-peptide chemistries, this approach is not available. The central challenge in this field is thus the following question: how can we identify useful sequences from among the multitude of sequences available to a sequence-controlled copolymer?

Over the past several years, the field has begun applying AI tools, including neural networks, to this problem, with some success. A number of AI-based approaches have aimed to accelerate solution of the \emph{forward} problem of predicting properties from sequence. For example, unsupervised learning methods have been employed to classify and predict sequence-mediated assembled structures formed by sequence-controlled copolymers.\cite{statt_unsupervised_2021, bhattacharya_predicting_2022}. Supervised machine learning via Neural Networks has been employed to predict gyration radii of single-chain particles based on their backbone sequence\cite{webb_targeted_2020, sramesh_polymer_2023}, the sequence-dependence of interfacial tension reduction by copolymer compatibilizers\cite{himanshu_when_2022}, and the sequence-dependence of the heat capacity and relaxation times \cite{patel_featurization_2022}. 

However, a central challenge is as follows. Because these systems, as noted above, do not possess significant databases of pre-existing data, it is necessary to generate large datasets for each problem prior to applying these AI tools. As the primary methods for generating datasets for sequence-controlled polymers - experiments and molecular dynamics simulations - are both relatively slow, the capacity of these AI tools to enable prediction and design remains sharply limited compared to their analogues for natural biological macromolecules. For these reasons, studies, such as those discussed, that employ AI to probe purely synthetic sequence-controlled macromolecules above have generally been limited to simple bead-based generic polymer models. Given that these data generation tools are unlikely to dramatically increase in throughput in the near future, a breakthrough in the capacity to make AI-based predictions based upon more limited datasets is required - particularly if an extension to design of fully detailed real-world chemistries is to be realized.

Here, we particularly target two opportunities to dramatically reduce the amount of data required for AI to accelerate the design of a molecular sequence.

\begin{enumerate}
    \item At present, effective use of AI requires extensive datasets at each particular set of conditions or chemistries of interest. For example, if one has conducted an AI-based design process for sequence effects at a particular combination of temperature, concentration, and chemistries under consideration, an entirely new (and equally large) dataset will be required to train a similar AI tool at a second set of these conditions. A capability enabling AI-based predictions with one set of conditions to bootstrap AI learning at a second set of conditions could dramatically reduce the total amount of data required to predict and design properties in previously unexplored sequence spaces.
    \item Molecular simulations that have the capacity to produce data for AI training in these spaces face a set of core tradeoffs between simulation time and data quality that currently sharply limits simulation-based data-generation campaigns. One such tradeoff is a simple exchange between noise and simulation time - noise in properties of interest can commonly be reduced with increased simulated systems sizes or increased simulation times (commonly scaling roughly as the inverse root of either quantity as per the central limit theorem), but with, in many cases, a nearly linear resulting increase in simulation times. At a deeper level, coarse-grained models can accelerate simulation by roughly 100-fold relative to fully atomistic models~\cite{hung_universal_2018}, but often at the cost of considerably reduced accuracy of results. An AI-based method for learning on coarse models or noisy data and then applying these lessons to accelerate learning on more accurate and precise data could dramatically reduce the amount of simulations needed at the higher-fidelity, thus much slower, level, thus dramatically reducing the total simulation time needed to enable design. 
\end{enumerate}

We describe a new AI approach to overcoming these challenges in sequence-controlled polymers - a few-shot learning approach, wherein a deep neural network (DNN) is initially trained on sequence-property relationships from high-noise data at thermodynamic condition I, and then is tuned on low-noise data from on sequence-property relations under a distinct thermodynamic condition II, where the sequence-property relationship is different. We show that this enables learning and prediction at condition II using far less data than would be required if a DNN were trained against space II from scratch.

 We specifically implement and test this advance within the space of \emph{sequence-controlled copolymer compatibilizers}. Macromolecular compatibilizers are of high interest for a range of applications, from enabling mixed-waste-stream post-consumer plastics recycling \cite{lin_advances_2024, zhou_understanding_2024, shen_threading--needle_2023, qian_design_2023, self_linear_2022} to fabrication of high-performance biopolymer composites \cite{fredi_compatibilization_2024, zhao_super_2020, zeng_compatibilization_2015,lyatskaya_designing_1996} to stabilization of emulsions. At the polymer-polymer interfaces characteristic of recycling and polymer composite applications, several sequence-dependent mechanisms play a role in compatibilization. First, compatibilizers reduce interfacial tension \cite{zhou_compatibilization_2021, meenakshisundaram_designing_2017,patterson_monomer_2020,zhou_understanding_2024}, which favors smaller and more intercalcated polymer domains yielding improved mechanical properties. Second, compatibilizers possessing long loops can in some cases entangle with polymer domains or thread the needs of loops protruding from crystalline regions in polymer domains, enhancing mechanical properties via a molecular `knitting' effect.\cite{lin_advances_2024, shen_threading--needle_2023,qian_design_2023,self_linear_2022} Finally, in some cases compatibilizers can co-crystallize with crystalline domains in semicrystalline host polymers.\cite{lin_advances_2024,fredi_compatibilization_2024, shen_threading--needle_2023,qian_design_2023,self_linear_2022} Here we focus on simulations probing sequence effects on the first of these mechanisms. This is of both practical and fundamental interest: interfacial tension $\gamma$ reduction, entanglement and cocrystallization mechanisms appear to possess distinct dependences on segment sequence. This both challenges the design of compatibilizer sequences integrating multiple reinforcement mechanisms and leaves open the foundational question of the mechanisms of this sequence control.

Our prior work has demonstrated, in bead spring model simulations, that the surface-tension-reduction contribution to compatibilization is highly sensitive to sequence, with non-trivial Genetic-Algorithm(GA)\cite{paszkowicz_genetic_2009,chakraborti_genetic_2004}-designed sequences exhibiting interfacial energy reductions considerably exceeeding the best results obtained by `intuitive' blocky polymers\cite{meenakshisundaram_designing_2017}. The optimal sequence is highly sensitive even to factors such as compatibilizer concentration; it is expected to likewise be quite sensitive to  system chemistry and the innate interfacial tension of the target system. At present, AI-based design requires a dataset for each particular set of conditions of relevance (e.g. each distinct compatibilizer concentration, system chemistry, temperature, and other such variables). Design of these systems would be dramatically accelerated if data from one set of conditions could be employed to significantly accelerate AI learning under a distinct set of conditions.

Using the approach outlined above, here we address this challenge by showing that good predictions can be made on this basis with much smaller datasets than would be required were the DNN trained on set B without prior priming. This result suggests a new approach wherein an AI tool is trained on relatively large, noisy datasets at one `priming' thermodynamic condition, and then tuned on smaller, higher fidelity datasets for other conditions. This could dramatically reduce the amount of data that is necessary to obtain over a range of chemical and thermodynamic conditions to make good predictions based on AI.

\section{Materials and Methods}

\subsection*{Molecular Simulations and Resulting Datasets}
We employ data on the simulated interfacial energy reduction obtained by introducing sequence-controlled bead-spring compatibilizers to a bead-spring polymer-polymer interface, generated in our prior work.\cite{meenakshisundaram_designing_2017} In summary, the simulated system is as follows. Our simulations consist of two layers of highly incompatible polymers. Polymer chains are modeled via an attractive variant of the classical Kremer-Grest polymer model \cite{grest_molecular-dynamics_1986}. Each chain is comprised of 20 beads, interacting via a Lennard Jones potential,

\begin{equation} \label{eq:binlj}
{{E}_{ij}}=4{{\varepsilon }\left[ {{\left( \frac{\sigma }{r} \right)}^{12}}-{{\left( \frac{\sigma }{r} \right)}^{6}} \right], r<r_{cut} \quad}
\end{equation}
where $i$ and $j$ denote the interacting bead types, r is the distance between beads, $\varepsilon=1$ is an interaction strength parameter, $\sigma=1$ is a range parameter, and the interaction is cut off at a distance $r_{cut}$. Each polymer consists of a distinct bead type (A or B). The bead types' self-interactions are the same, such that the two polymers are identical in their pure-state parameters. However, they are highly immiscible, which is implemented by truncating their cross-interaction (A-B) potential at the minimum of the pair potential, i.e. $r_{cut}=2^{1/6}\sigma$. This potential and its variants have been widely employed to model polymer interfacial phenomena \cite{estridge_diblock_2015, seo_effect_2015, baschnagel_computer_2005, ghanekarade_combined_2022, tulsi_hierarchical_2022}.

\begin{figure}[!h]
\includegraphics[width=0.9\linewidth]{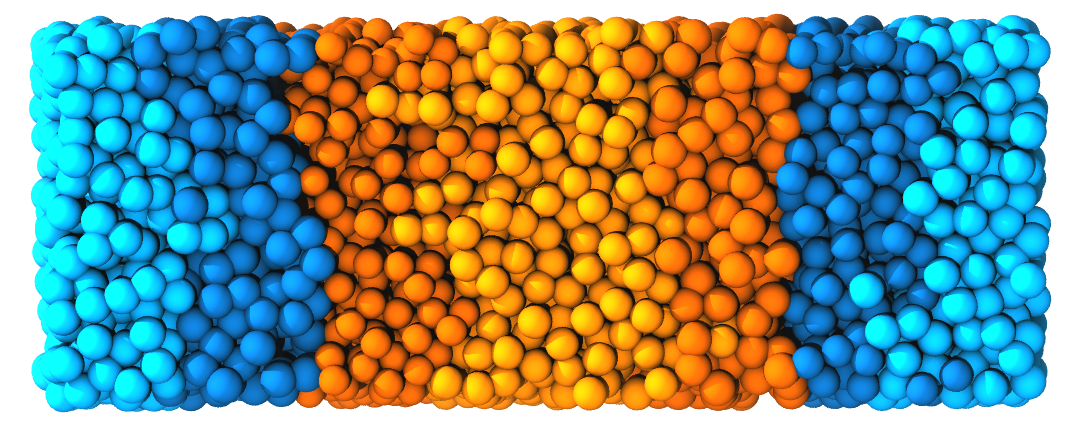}
\caption{{\bf  }Rendering (performed with VMD\cite{humprey_vmd_1996}) of a representative simulated system containing a diblock compatibilizer. The two homopolymers are rendered in light orange and blue, with the corresponding compatibilizer groups shown in darker orange and blue.}
\label{fig:simulationrender}
\end{figure}

Copolymer compatibilizer chains are constructed from the same bead types (A and B) that comprise the two immiscible polymers, but covalently linked within a single chain, such that each compatibilizer consists of a unique A-B sequence. All compatibilizers in a given simulation are identical - i.e. they are sequence monodisperse. Compatibilizers are introduced in equal numbers at each interface at the target areal concentration, and the system is subject to a long isothermal equilibration period prior to data collection.  As shown in Fig. \ref{fig:simulationrender}, this leads to formation of a phase-separated structure, within which compatibilizer molecules localize to the interfaces. 

An interfacial energy between phases is then computed from the system's pressure anisotropy via the following standard equation \cite{harris_liquid-vapor_1992}:

\begin{equation} \label{eq:gamma}
{{\gamma }_{12}}=\left\langle \frac{{{L}_{z}}}{2}\left( {{P}_{zz}}-\frac{1}{2}\left( {{P}_{xx}}+{{P}_{yy}} \right) \right) \right\rangle 
\end{equation}

Here $L_z$ is the length of the simulation box in the z direction, and  $P_{zz}, P_{xx}, and P_{yy}$ are are the pressures acting in the x, y and z directions.

The system possesses an interfacial energy of 1.84 in reduced Lennard Jones (LJ) units prior to introduction of a compatibilizer.\cite{meenakshisundaram_designing_2017} In our prior study, we generated large datasets of sequence-interfacial energy relationships by selecting compatibilizer sequences for simulation via a genetic algorithm that optimized the system for low interfacial energy (large interfacial energy reduction via compatibilization). This algorithm was separately run at multiple target areal concentrations of compatibilizers, yielding distinct datasets (and distinct optimized compatibilizer sequences) at each concentration. These simulations are subject to the tradeoff between simulation time and signal-to-noise ratio described in the introduction. For a given areal concentration for compatibilizer molecules, a larger interfacial area will yield improved interfacial tension statistics, at the cost of increased total time that must be spent on simulation. 

In implementing and assessing a DNN strategy for knowledge transference between distinct conditions, we employ two datasets from a study.\cite{meenakshisundaram_designing_2017} These datasets differ both in the \emph{compatibilizer concentration} and in the \emph{simulated interfacial area (and thus data quality and simulation rate)}, in order to test the ability to overcome both challenges (1) and (2) described in the introduction. 

\textbf{Dataset I} is obtained from a set of simulations performed at a cross-sectional area of $12 \sigma \times 12\sigma $, at a concentration of 0.0833 per $\sigma^2$. 

\textbf{Dataset II} corresponds to simulations performed at 9 times the cross-sectional area, i.e. $36 \sigma \times 36\sigma $ (thus approximately 3 times less noise in the data), and at a distinct concentration of 0.0694 compatibilizer chains per $\sigma^2$. 

Dataset I, due to its much smaller simulated systems size (9 times fewer particles), contains more distinct sequence-interfacial-tension datapoints - 3232. Dataset II contains data from 1632 simulations. The data from each simulation consist of a binary sequence denoting the order of repeat units of type matching each incompatible polymer, together with the associated simulated interfacial tension of the system.

\subsection*{Machine Learning}
\begin{figure}[!h]
\includegraphics[width=0.9\linewidth, height=10cm]{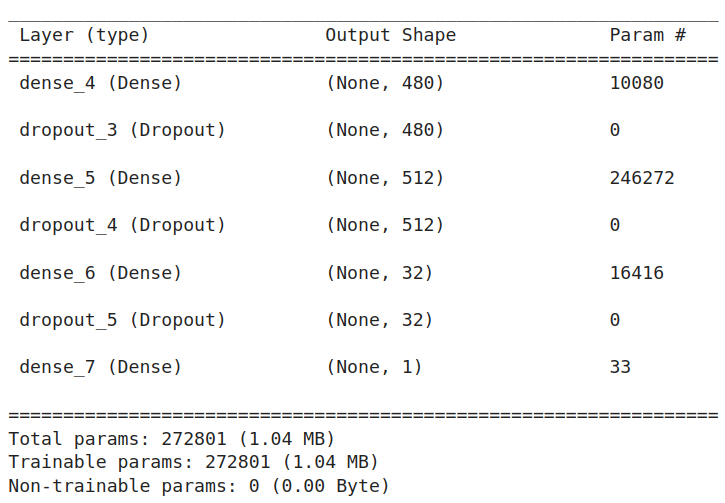}
\caption{{\bf  }Proposed Neural Network}
\label{model}
\end{figure}

We employed a multi-hidden layer (deep) neural network (DNN) to machine learn the relationships between sequence and interfacial tension in these systems. To generate an effective neural network architecture in this system, we employed AutoKeras.\cite{AutoKeras}. The resulting DNN is structured as follows (see Fig~\ref{model}). It begins with an input layer that accepts 20 features, with each feature specifying the identity of a repeat unit (0 for A, 1 for B, in sequence) within the 20-repeat-unit compatibilizer chains. The model includes three hidden layers: the first hidden layer has 480 units with ReLU (Rectified Linear Unit) activation and uses the RandomNormal initializer, while the second hidden layer consists of 512 units and also uses ReLU activation with the same initialization. The third hidden layer has 32 units with ReLU activation. The output layer contains a single unit with a linear activation function, appropriate for regression. The weights are initialized using a RandomNormal distribution with a mean of 0.0 and a standard deviation of 0.05. To reduce overfitting, a dropout rate of 0.2 was applied after each of the hidden layers. The model uses the AdamW optimizer, a learning rate of 0.0001, and uses mean squared error (MSE) as the loss function. The model was trained using up to 2000 epochs, with early stopping applied if the validation loss did not improve for 200 consecutive epochs. We used a batch size of 16, and selected the final model based on the lowest validation loss.

Through these steps, we aimed to explore the models' initial performance on individual datasets, its ability to generalize across datasets without retraining, and the impact of gradually combining examples from a new dataset on their prediction accuracy.


\section{Results}

\subsection{Concentration-dependence of coarse sequence-property relationship}

\begin{figure}[!h]
\includegraphics[width=0.9\linewidth]{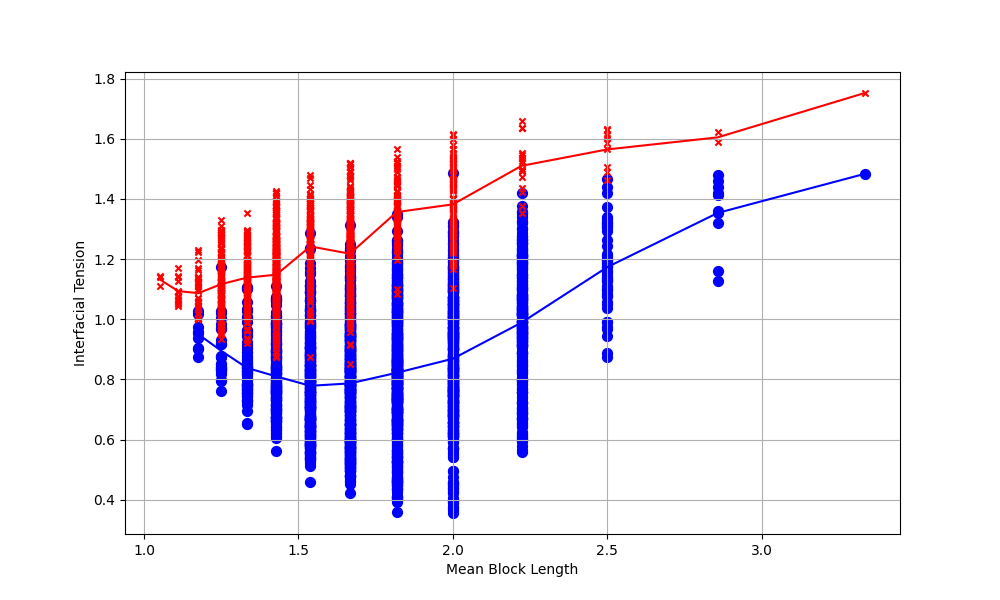}
\caption{{\bf  }Interfacial tension vs mean block length for each simulated system in dataset I (areal concentration $0.0833/\sigma^2$, blue circles), and dataset II (areal concentration $0.0694/\sigma^2$, red x's, and trend curves reporting on the average interfacial tension for candidates at each mean block length in corresponding colored lines. Results indicate that the two datasets, even at this coarse level of sequence representation, have quite distinct sequence-interfacial tension relationships.}
\label{fig:interfacialtension}
\end{figure}

Crucially, the sequence-interfacial-tension relationship is quite sensitive to concentration, such that the two datasets employed in this system involve both different levels of fidelity (higher noise in I, lower noise in II), and distinct sequence-interfacial-tension relationships. This can be seen in Fig~\ref{fig:interfacialtension}, which demonstrates that the dependence of interfacial tension on mean block size (i.e. the mean run length of type 1 or 2 repeat units - a very coarse representation of sequence) is quite different in the two cases. This indicates that any effort to tune a DNN, initially trained on dataset I, on dataset II (or vice versa) likely requires a nontrivial relearning of the sequence/interfacial tension relationship. The central AI task is thus to, first, learn the relationship between molecular sequence and interfacial tension at condition I, and then employ some of the knowledge gained in this process to reduce the amount of data required to learn the sequence/tension relationship at condition II.

\subsection{Single-dataset learning}

We initially employed the DNN to learn sequence-$\gamma$ relationships in each individual dataset, to assess the DNN's performance on the single-dataset level. To test for reproducibility, we randomly shuffled each dataset using 10 different seeds, ensuring variability in the training, validation, and testing subsets. Training data was 80\% of the overall data, validation was a disjoint 10\% of the data and testing was another disjoint 10\% of the data. For each split, model performance was assessed on the test set using two standard regression metrics: the coefficient of determination ($R^2$) and root mean squared error (RMSE). The results are summarized in Tables~\ref{table1} and Table~\ref{table2} for Datasets I and II, respectively. Notably, the mean $R^2$ is appreciably higher and and RMSE appreciably lower for dataset II than I, reflecting the noisier character of dataset I, a result of the smaller simulated system size of the underlying simulations as discussed above.

Despite this, in both cases the DNN provides effective prediction of sequence-$\gamma$ relationships, as indicated by several measures of predictive power. First, the $R^2$ values in both cases are of order 0.9, which is comparable to some of the best $R^2$ values obtained for similar sequence-$\gamma$ prediction efforts  \cite{himanshu_developing_2023}. Second, and more practically, the RMSE values are quite small as compared to the range of values of interest for design. As discussed above, the neat (uncompatibilized) interface has an interfacial tension $\gamma$ of 1.84. In general, the physical goal in such systems is to reduce $\gamma$ as close to zero as possible to encourage extensive and durable intercalcation of the domains under mixing. The total range of $\gamma$ values of interest is thus 0 to 1.84. The RMSE values of 0.065 (dataset I) and 0.039 (dataset II) represent relative RMSEs of 3.5\% and 2.1\% of the value range of interest, indicating a high degree of capability to discriminate between `good' and `bad' compatibilizer sequences - the basic criterion for enabling an effective sequence-screening process.

In addition, we also analyzed how well the model could recover the sequences with the lowest interfacial tension. Specifically, we compared the 20 simulated sequences with the lowest true $\gamma$ values to the 20 sequences with lowest $\gamma$ values predicted by the model for each seed, and calculated the number of matches. This top-20 match count serves as a practical measure of the model’s ability to prioritize top candidate compatibilizers. DNNs trained on dataset II consistently exhibited a higher number of correct top-20 predictions; this is again a consequence of less noise further underscoring its predictive reliability. Detailed results of top-20 match counts for each datasets is provided in Table~\ref{tab:top20_matches}.

\begin{table}[!ht]
\centering
\caption{
{\bf Evaluation of $R^2$ and RMSE on Dataset I Across Different Data Splits for Training(80\%), Validation(10\%), and Testing(10\%)}}
\begin{tabular}{|l|l|l|l|}
\hline
\multicolumn{4}{|c|}{\bf $R^2$ and RMSE on Dataset I} \\ \hline
\textbf{Split} & \textbf{Seed Value} & \textbf{$R^2$} & \textbf{RMSE}  \\ \hline
Split 1 & 5 & 0.884 & 0.069  \\ \hline
Split 2 & 17 & 0.880 & 0.065 \\ \hline
Split 3 & 29 & 0.898 & 0.065 \\ \hline
Split 4 & 43 & 0.898 & 0.059 \\ \hline
Split 5 & 53 & 0.897 & 0.059 \\ \hline
Split 6 & 311 & 0.890 & 0.068 \\ \hline
Split 7 & 331 & 0.898 & 0.065 \\ \hline
Split 8 & 631 & 0.876 & 0.067 \\ \hline
Split 9 & 719 & 0.877 & 0.067 \\ \hline
Split 10 & 769 & 0.868 & 0.070 \\ \hline
&Average  & 0.887 & 0.065 \\ \hline
&Standard Deviation  & 0.011 & 0.004 \\ \hline
\end{tabular}
\label{table1}
\end{table}


\begin{table}[!ht]
\centering
\caption{
{\bf Evaluation of $R^2$ and RMSE on Dataset II Across Different Data Splits for Training(80\%), Validation(10\%), and Testing(10\%)}}
\begin{tabular}{|l|l|l|l|}
\hline
\multicolumn{4}{|c|}{\bf $R^2$ and RMSE on Dataset II} \\ \hline
\textbf{Split} & \textbf{Seed Value} & \textbf{$R^2$} & \textbf{RMSE}  \\ \hline
Split 1 & 17 & 0.911 & 0.038 \\ \hline
Split 2 & 19 & 0.902 & 0.041 \\ \hline
Split 3 & 29 & 0.920 & 0.040 \\ \hline
Split 4 & 73 & 0.932 & 0.035 \\ \hline
Split 5 & 269 & 0.907 & 0.0387 \\ \hline
Split 6 & 307 & 0.937 & 0.037 \\ \hline
Split 7 & 389 & 0.872 & 0.047 \\ \hline
Split 8 & 433 & 0.923 & 0.037 \\ \hline
Split 9 & 509 & 0.914 & 0.039 \\ \hline
Split 10 & 647 & 0.922 & 0.039 \\ \hline
&Average  & 0.914 & 0.039 \\ \hline
&Standard Deviation  & 0.017 & 0.003 \\  \hline
\end{tabular}
\label{table2}
\end{table}

\begin{table}[!ht]
\centering
\caption{{\bf Comparison of Top-20 Sequence Match Counts Across Splits for Dataset I and II}}
\label{tab:top20_matches}
\begin{tabular}{|l|c|c|c|}
\hline
\textbf{Split} & \textbf{Seed Value} & \textbf{Dataset I} & \textbf{Dataset II} \\ \hline
Split 1 & 5   & 14 & 16 \\ \hline
Split 2 & 17  & 14 & 15 \\ \hline
Split 3 & 29  & 13 & 17 \\ \hline
Split 4 & 43  & 12 & 17 \\ \hline
Split 5 & 53  & 13 & 15 \\ \hline
Split 6 & 311 & 12 & 17 \\ \hline
Split 7 & 331 & 16 & 14 \\ \hline
Split 8 & 631 & 15 & 16 \\ \hline
Split 9 & 719 & 9  & 17 \\ \hline
Split 10 & 769 & 14 & 16 \\ \hline
{\bf Average} & & 13.2&  16 \\ \hline
\end{tabular}
\begin{flushleft}
\small The table shows the number of overlapping sequences between the top 20 predicted and actual lowest-$\gamma$ sequences for each data split. Higher match counts in Dataset II reflect stronger predictive ranking performance.
\end{flushleft}
\end{table}

\subsection{Zero-shot and Few-shot Learning}

Given that the DNN provides strong predictive performance for sequence–$\gamma$ relationships within individual datasets, we next examined whether this predictive capability could generalize across datasets collected under different thermodynamic conditions. In particular, we investigated whether models trained on one dataset could be leveraged to accelerate learning on a second, related dataset—thus testing the potential for cross-system knowledge transfer.

As a baseline, we first evaluated zero-shot performance, where models trained on Dataset I were directly tested on Dataset II, and vice versa. As expected, the performance was modest due to the difference in the underlying sequence-$\gamma$ relationships in these two datasets (Figure~\ref{fig:interfacialtension}), which are obtained at different compatibilizer concentrations. These results establish that direct transfer without tuning is not effective and motivates the need for adaptation.

We then explored few-shot transfer by fine-tuning models trained on Dataset I using incremental amounts—specifically $5\%$, $10\%$, $20\%$, $30\%$, $40\%$, $50\%$, $60\%$, $70\%$, and $80\%$—of data from Dataset II. The data from Dataset II was randomly shuffled once at the beginning with seed 42 and used consistently across all fine-tuning increments. For each increment, 10 models were initialized using weights from the 10 independently trained models on Dataset I (each trained with a different seed). Each of these models was fine-tuned on the specified portion of Dataset II, validated on the final $10\%$, and tested on the remaining data (excluding both the fine-tuning and validation subsets). Fine-tuning was performed for up to 1000 epochs, with early stopping applied if the validation loss did not improve for 100 consecutive epochs. Training was performed using a batch size of 16, a learning rate of 1e-4, and the Adam optimizer, with mean squared error (MSE) as the loss function. For each fine-tuning increment, the same training, validation, and test splits were used across all 10 models, each initialized from a different pre-trained seed model. We then averaged the resulting $R^2$ and RMSE scores. The same procedure was applied in the reverse transfer direction, wherein models pre-trained on Dataset II were fine-tuned using subsets of Dataset~I.

\begin{figure}[H]
    \centering
    \begin{subfigure}{0.48\textwidth}
        \centering
        \includegraphics[width=0.9\linewidth]{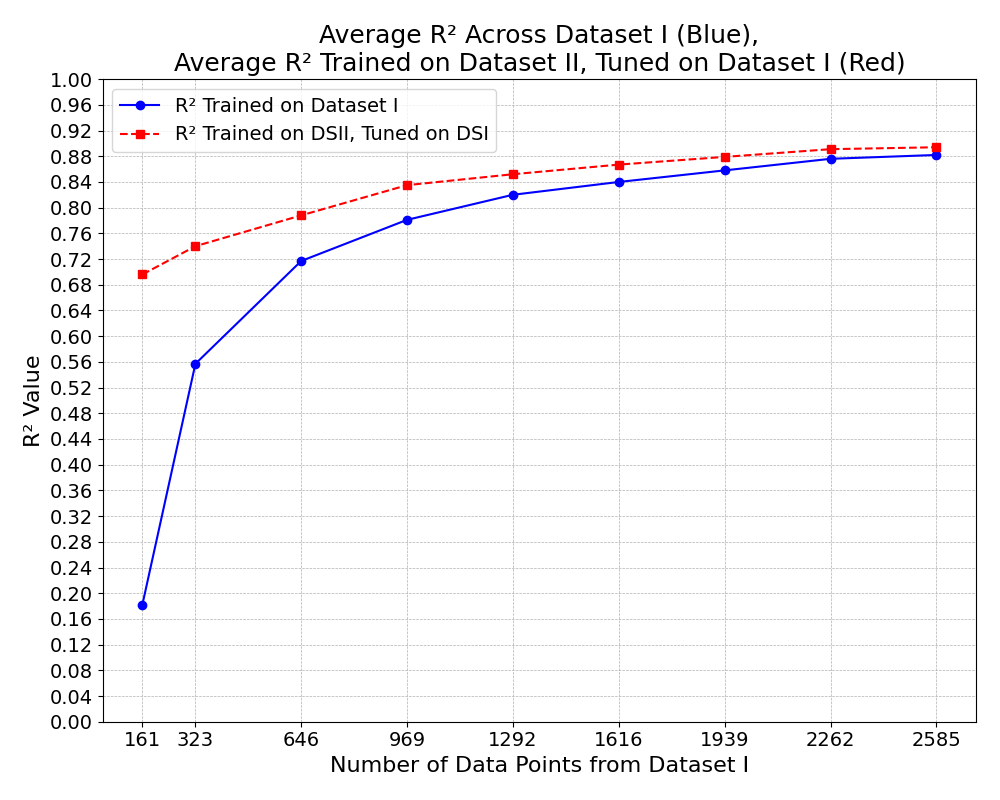}
 \caption{}
        \label{fig:subim21}
    \end{subfigure}
    \hfill
    \begin{subfigure}{0.48\textwidth}
        \centering
        \includegraphics[width=0.9\linewidth]{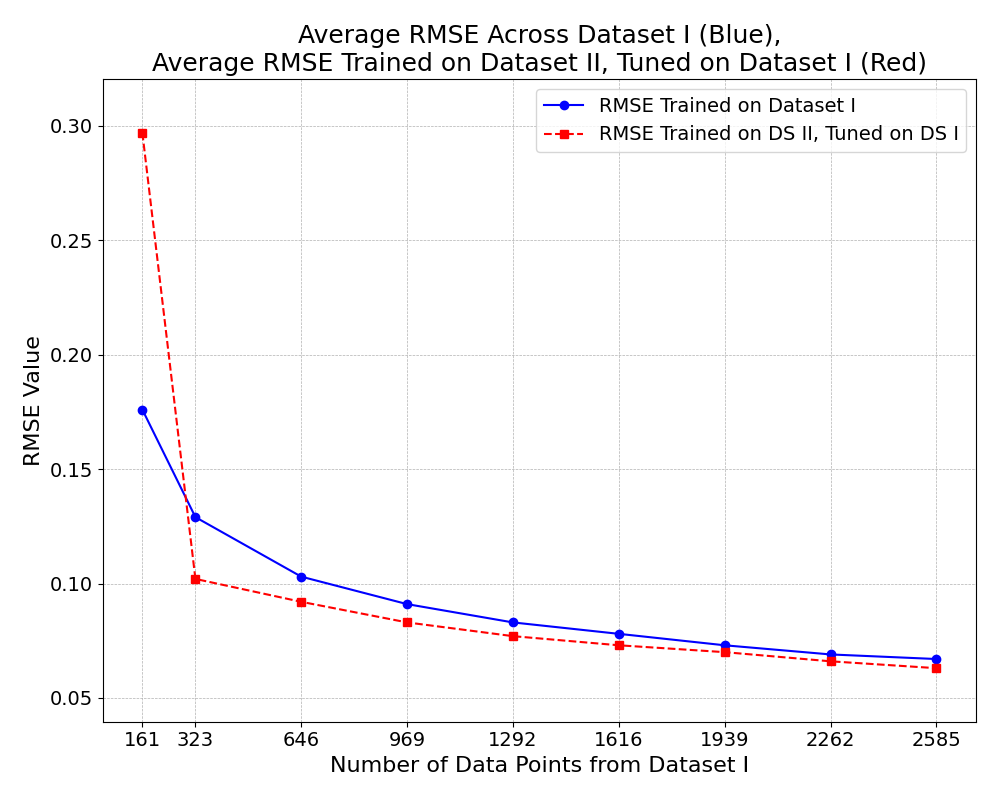}
 \caption{}
        \label{fig:subim21_rmse}
    \end{subfigure}

    \vspace{0.5cm}
    \caption{Few-shot tuning results: Models trained on Dataset II and incrementally tuned using Dataset I, as well as models trained just on Dataset I with different training set sizes. a) Average $R^2$ across 10 models, b) Average RMSE values across 10 models.  Model performance improves with the amount of target data introduced and tuning does better than training on all data.}
    \label{fig:fig1}
\end{figure}

The results, shown in Figure~\ref{fig:fig1}, demonstrate that as more target-domain data (from dataset II) is introduced, $R^2$ increases and the RMSE drops. As a baseline comparison for measuring the value of `priming' on dataset I prior to tuning on dataset II we compare to $R^2$ and RMSE values for training \emph{directly} on dataset II. As can be seen in Figure~\ref{fig:fig1}, the DNN primed on dataset I is able to learn sequence-property relationships in dataset II far more rapidly than a de novo DNN. As one point of comparison, the DNN primed on dataset I attains an $R^2$ value of 0.8 with less than a quarter of the training data as the de novo model trained on dataset II.


This indicates that even small amounts of additional data allow the pre-trained model to adapt effectively.

\begin{figure}[H]
    \centering
    \begin{subfigure}{0.48\textwidth}
        \centering
        \includegraphics[width=0.9\linewidth]{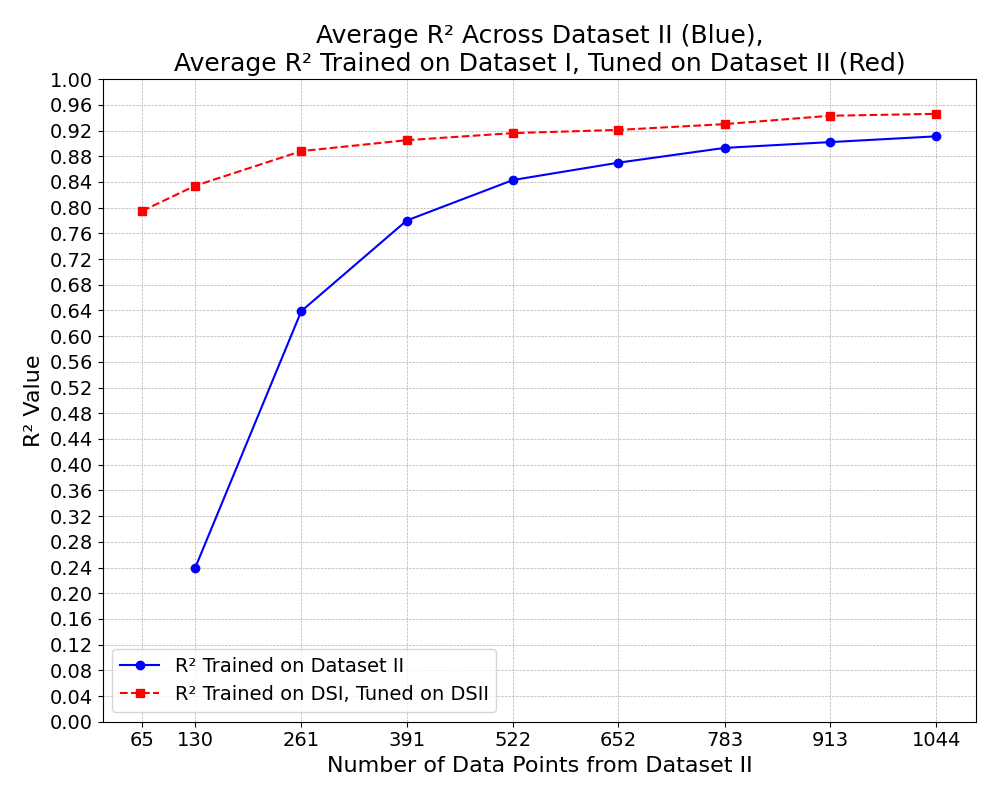}
\caption{}
    \end{subfigure}
    \hfill
    \begin{subfigure}{0.48\textwidth}
        \centering
        \includegraphics[width=0.9\linewidth]{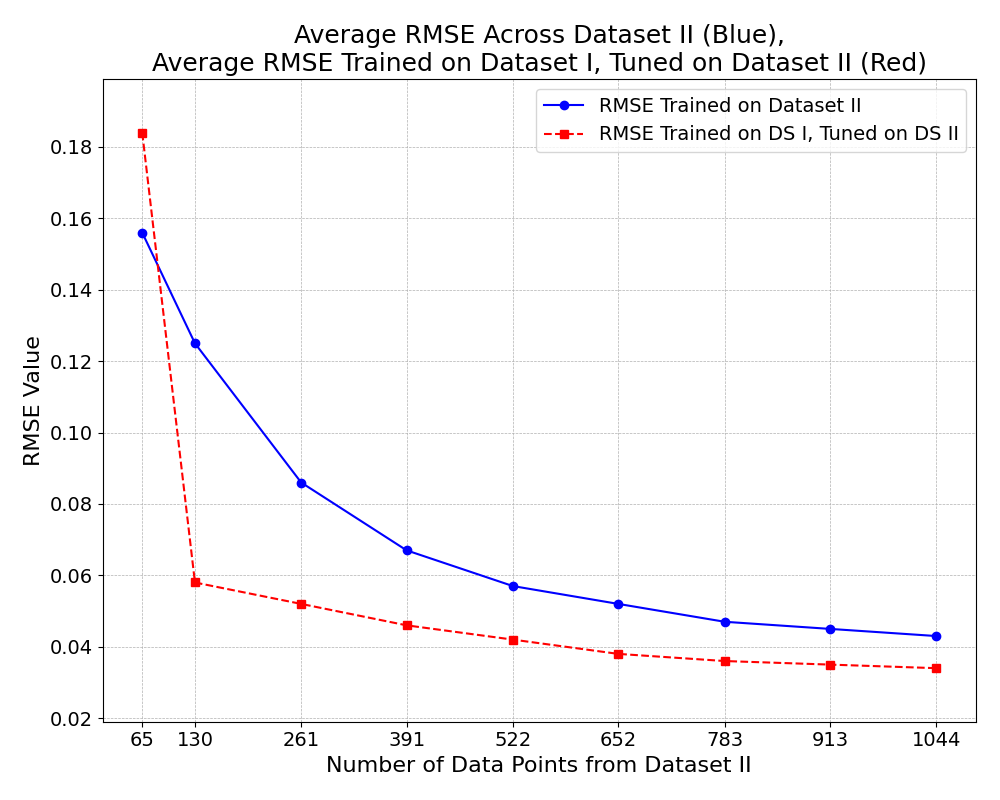}
 \caption{}
        \label{fig:subim22_rmse}
    \end{subfigure}

    \vspace{0.5cm}
    \caption{Few-shot tuning results: Models trained on Dataset I and incrementally tuned using Dataset II, as well as models trained just on Dataset II with different training set sizes. a) Average $R^2$ across 10 models, b) Average RMSE values across 10 models. Model performance improves with the amount of target data introduced and tuning does better than training on all data.}
    \label{fig:fig2}
\end{figure}

We then reversed the transfer direction. Models initially trained on dataset II were fine-tuned with incremental amounts of data from dataset I. The same procedure was followed for each tuning level, again the last 10\% of dataset I was left for validation, and the remainder was used for testing. The results, again averaged across 10 random seed splits, are shown in Figure~\ref{fig:fig2}.

As seen in Figure~\ref{fig:fig2}, the trend remains consistent. The performance steadily improves as more tuning data are introduced. 

The materials with the lowest interfacial tension are most interesting.  Hence, an experiment was done to see how many of the lowest interfacial tensions predicted were in the ground truth set of lowest values.  This was done with zero and few shot learning.  To assess the robustness of model predictions under few-shot tuning, we evaluated the standard error of identifying the lowest twenty interfacial tension ($\gamma$) sequences in the test dataset. Specifically, we aimed to assess how consistently each model, after tuning, could identify the true lowest-$\gamma$ instances across multiple random initializations.

We used 10 models trained on different splits of  Dataset I and fine-tuned them using increasing portions of  Dataset II—from $5\%$ to $80\%$. After each tuning step, the number of correctly identified sequences among the twenty lowest-$\gamma$ instances in the test set was recorded. The standard error at each tuning level was computed as:
\[
\text{SE} = \frac{\sigma}{\sqrt{n}},
\]
where $\sigma$ is the standard deviation of the prediction counts across the 10 models, and $n = 10$ is the number of splits. A lower standard error implies more stable and reliable performance across different random initializations.

\begin{figure}[H]
    \centering
    \begin{subfigure}{0.48\textwidth}
        \centering
        \includegraphics[width=0.9\linewidth]{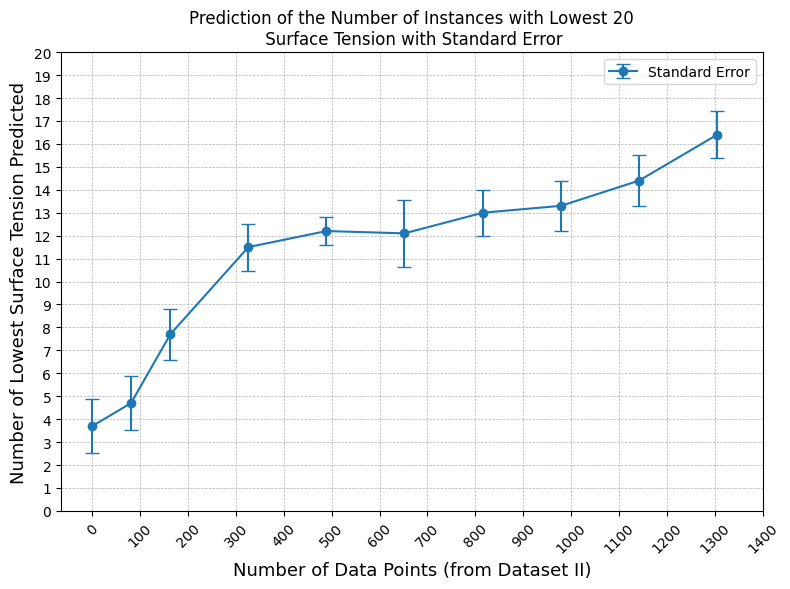}
        \caption{Standard error of top-20 match counts across 10 splits: Dataset I $\rightarrow$ Dataset II.}
        \label{fig:subim31}
    \end{subfigure}
    \hfill
    \begin{subfigure}{0.48\textwidth}
        \centering
        \includegraphics[width=0.9\linewidth]{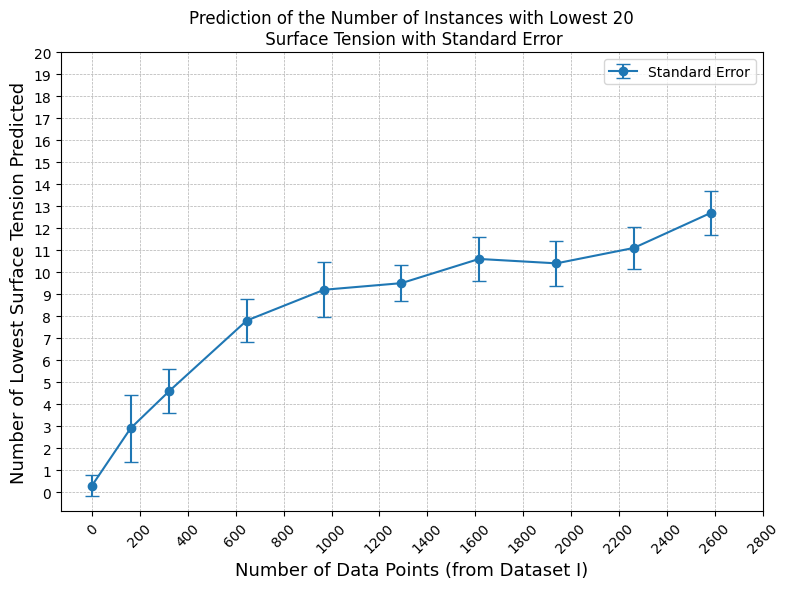}
        \caption{Standard error of top-20 match counts: Dataset II $\rightarrow$ Dataset I.}
        \label{fig:subim32}
    \end{subfigure}

    \caption{Standard error in predicting the lowest-$\gamma$ sequences during transfer learning. Lower values reflect more stable predictions across random splits.}
    \label{fig:image3}
\end{figure}

This same procedure was mirrored in the reverse transfer setting, where models trained on the Dataset II were fine-tuned using increasing portions of  Dataset I. The standard error values from both directions are shown in Figure~\ref{fig:image3}.  It can be seen that with as little as 15\% of data for tuning between 4 and 7 of the lowest surface tension ground truth examples are in the lowest 20 $R^2$ predictions when tuning Dataset I or Dataset II.  With 20\% of the data this improves to between 8 and 11. 

\section{Conclusion}

Design of sequence-controlled synthetic polymers remains a generational challenge. These systems lack the massive evolved datasets of biological proteins that can accelerate design of bio-modified proteins. They demand new approaches to accelerated design in the absence of large sets of pre-existing data. Here we demonstrate a new approach to AI-accellerated design of polymer sequence with far less data. In this approach, a deep neural network is trained on coarse sequence-property relationships obtained from high-noise data at one set of conditions and can then be readily tuned to make higher-fidelity predictions at a distinct set of conditions, associated with distinct sequence-property relations. Crucially, we show that tuning on the second dataset requires far less data to make good predictions than would be required without prior priming. 

This priming-and-tuning approach opens the way to using a single large parent dataset to dramatically accelerate prediction and design in an entire constellation of related systems. Deep Neural Networks, like most AI, generally struggle to extrapolate from their training space to even reasonably closely related spaces outside the training set.~\cite{noauthor_national_2021,noauthor_national_2023,zhang_deep_2021,zhang_understanding_2021} This approach provides a relatively data-cheap means of enabling deep neural networks to translate prior insights to new related systems. This could greatly reduce the amount of new data necessary to design sequence-controlled polymers and other similar chemical systems. 

Finally, this method's ability to translate from high-noise to low-noise data hints at new opportunities to employ deep neural network priming-and-tuning for accelerated molecular prediction. When aiming to quickly gain insights into a new molecular system, humans routinely employ  coarse-grained molecular simulations, which aim to capture coarse features of the system quickly but lack the full chemical fidelity to make quantitative predictions. Mapping insights from such simple models to real systems typically requires human intuition, development of master equations, and considerable time. The approach developed here, with further augmentation, may open the way to priming neural networks on coarse models and then tuning them on data from quantitative atomistic simulations. This could enable deep neural networks to translate insights from coarse to quantitative models, acclerating (or even replacing) this key step in efficient materials prediciton and design.


\bibliography{achemso-demo}

\end{document}